# Quantum random number generators and their use in cryptography

M. Stipčević*
* University of Santa Barbara, Santa Barbara, California, USA, and
Rudjer Bošković Institute, Division of Experimental Physics, Zagreb, Croatia
stipcevi@gmail.com

*Abstract* - Random number generators (RNG) are an important resource in many areas: cryptography (both quantum and classical), probabilistic computation (Monte Carlo methods), numerical simulations, industrial testing and labeling, hazard games, scientific research, etc. Because today's computers are deterministic, they can not create random numbers unless complemented with a RNG. Randomness of a RNG can be precisely, scientifically characterized and measured. Especially valuable is the information-theoretic provable RNG (True RNG – TRNG) which, at state of the art, seem to be possible only by use of physical randomness inherent to certain (simple) quantum systems. On the other hand, current industry standard dictates use of RNG's based on free running oscillators (FRO) whose randomness is derived from electronics noise present in logic circuits and which cannot be strictly proven. This approach is currently used in 3-rd and 4-th generation FPGA and ASIC hardware, unsuitable for realization of quantum TRNG. We compare weak and strong aspects of the two approaches and discuss possibility of building quantum TRNG in the recently appeared Mixed Signal FPGA technology. Finally, we discuss several examples where use of a TRNG is critical and show how it can significantly improve security of cryptographic systems.

## I. INTRODUCTION

True random numbers, or more precisely nondeterministic random number generators, seem to be of an ever increasing importance. Random numbers are essential in cryptography (classical, stochastic and quantum), Monte Carlo calculations, numerical simulations, statistical research, randomized algorithms, lottery etc. Today, true random numbers are probably most critically required in cryptography and its numerous applications to cyber-security such as: Smart Energy Grid, e-banking, internet trade, prepaid cards etc.

In applications where provability is essential, randomness sources (if involved) must also be provably random otherwise the whole chain of proofs will collapse. In cryptography, where due to the Kerhoff's principle all parts of protocols are publicly known except some secret (the key or other information) known only to the sender and the recipient, it is clear that secret must not be calculable by an eavesdropper i.e. it must be *random*. For example the famous BB84 quantum key distribution protocol [1] (described below) would be completely insecure if only an eavesdropper could calculate (or predict) either Alice's random numbers or Bob's random numbers or both. From analysis of the secret key rate presented therein it is obvious that any guessability of random numbers by the eavesdropper would leak relevant information to him, thus diminishing the effective key rate. It is intriguing (and obvious) that in the case that the eavesdropper could calculate the numbers exactly, the cryptographic potential of the BB84 protocol would be zero. This example shows that the local random number generators assumed in BB84 are essential for its security and should not be taken for granted.

## II. RANDOM NUMBER GENERATION

Due to the Kerhoff's principle, the definition of a random number generator suitable for cryptography must include that even if everything is known about the generator (schematic, algorithms etc.) it still must produce totally unpredictable bits.

Historically, there are two approaches to random number generation: algorithmic (pseudorandom) and by hardware (nondeterministic). Without loss of generality in the rest of the article we will assume that generators produce random bits (binary values 0 or 1).

Pseudorandom number generators (PRNG) are well known in the art and we are not going to address them here. A PRNG is nothing more than a mathematical formula which produces deterministic, periodic sequence of numbers which is completely determined by the initial state called *seed*. By definition such generators are not provably random. In practice, PRNG's feature perfect balance between 0's and 1's (zero bias) but also strong long-range correlations which undermine cryptographic strength and can show up as unexpected errors in Monte Carlo calculations and modeling.

Advantages of PRNG's are their low cost, ease of implementation and user friendliness, especially in a CPU-available environment such as a PC computer.

In contrast to that, hardware random number generators (RNG for short) extract randomness form physical processes that behave in a nondeterministic way which makes them better candidates for true random number generation. Examples of such physical processes are: Johnson's noise, Zener noise, radioactive decay, photon path splitting at the two-way beam splitter, etc [2]-[6]. Unlike the PRNG's hardware generators suffer from bias *b*, defined as the difference of probabilities of 1's and 0's:

$$b = p(1) - p(0) \qquad (1)$$

and long-range correlations. Our definition of PRNG is not to be confused with PRNG's implemented in hardware.

*A. Noise based RNG's*

Johnson's effect ideally creates random voltage on terminals of any resistive material which is held at a temperature higher than absolute zero. Johnson's noise is due to random thermal motion of the quantized electric charge (i.e. carriers). However, long-range carrier correlations in conductors cause correlations in movements of electric charges and therefore the resulting voltage is not completely random.

Zener noise (in semiconductor Zener diodes) is caused by tunneling of carriers through quantum barrier of ideally constant height and width. If current is sufficiently low, individual "jumps" of carriers through barrier will be seen as voltage peaks across the diode forming a pink noise of perfect randomness. However, Zener effect is never found well isolated in physical devices from other effects nor is the quantum barrier constant. Most of the fore mentioned processes in resistors and Zener diodes have some memory effect. This means that an instantaneous voltage across the device depends on voltages in the (near) past and this in turn leads to a correlation among random numbers extracted there from. The biggest problem here is that non-randomness of noise sources can not be well characterized, measured or even controlled during fabrication. Furthermore, noise sources produce rather tiny voltages that need to be strongly amplified before conversion to digital form which introduces further deviations from randomness due to the limited amplifier bandwidth and gain non-linearity.

The general idea of noise based hardware RNG is the following. The random voltage is sampled periodically and compared to a certain pre-defined threshold: if higher then „1" is generated, otherwise „0" is generated (Fig. 2). It is obvious that threshold can be set so that the probabilities of 1's and 0's are roughly the same. However, fine tuning of the threshold poses an insurmountable time-consuming problem and can never be done properly. For example, if tuning of bias to value of 0.1 requires 10 seconds, then tuning to 10 times lower value (0.01) would take 100 times longer (the required time scales as square of improvement ratio). On top of that there is a problem of stability: even the smallest drift of the mean value (for example due to temperature or supply voltage change) will create a large bias. Therefore some kind of post-processing is required (for example von Neumann or Peres de-biasing). This still does not solve the problem of correlations among bits which can even be enhanced by de-biasing procedures or changed from short-range ones to long-range ones (see Section D).

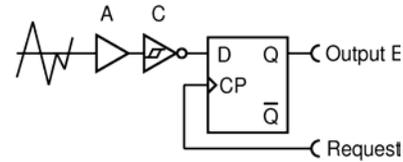

Fig. 2. Noise based RNG. A is an analog amplifier while C is a comparator whose output is either 0 or 1 depending whether its input is below or above certain adjustable threshold. Upon request, fresh new random bit will sit on the output Q.

In conclusion, a decent proof of randomness for present noise-based random number generators seems impossible because the underlying physical processes are not well isolated and proven to be random.

*B. Free runing oscillator RNG's*

When output of a logical inverter circuit is fed to its input, the circuit turns to an oscillator, so called *free running oscillator* (FRO), Fig 3.

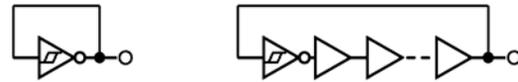

Fig. 3. Schematic diagram of a fast (left) and slow (right) FROs. Oscillation frequency is determined by internal delays and stray capacitances.

Frequency of oscillations is typically very sensitive to variation of power supply voltage and temperature. Noise present at the input causes very fast, random jitter of frequency and phase. Since noise of each such circuit is individual it is reasonable to assume the multiple oscillators even when on the same chip have different frequencies and that their mutual phases walk off randomly in time.

Basic principle of random number generation with FRO's is that that output of a fast FRO (which can be either logical 0 or logical 1) is sampled by a slow FRO. This is an equivalent of abrupt stopping of a quickly turning wheel of fortune. Because the wheel spins so "fast" it appears stopped at a "random" position. In case of two FRO's it is important that the relative phase jitter is random and large enough. Clearly, if there is no phase jitter the output will provide repetitive binary pattern.

Current industry standard [7] dictates use of RNG's based on free running oscillators (FRO) whose randomness is derived from electronics noise present in logic circuits. This approach is currently used in 3-rd and 4-th generation FPGA and ASIC hardware. As this infrastructure is unsuitable for realization of quantum RNG (see below), a FRO approach seem to be a reasonable viable alternative. However, semiconductor industry is making an enormous effort to make the noise as small as possible. Consequently the effect of jitter can be very small and cause the FRO based RNG to operate in nearly PRNG regime. Furthermore, present solutions described in the scientific journals and patents not only lack provability but in most cases even an attempt to prove randomness.

In conclusion, FRO based RNG's are low-cost, low-entropy solutions whose only good side is the fact that they can be easily implemented in conventional chips which are used in various cyber-security solutions.

*C. Quantum RNG's*

It turns out that some things in Nature come in the smallest amounts known as *quanta*. For example the electron carries the smallest quantity of charge, *e*. Similarly, there is the smallest quantity of information, called *qubit*. A single quantum of light (photon) can be used as a carrier of one qubit, but there are many other examples and they are not limited only to elementary particles. Qubit can be thought of as a linear combination of two bit values: 0 and 1. When a certain type of measurement is performed on a qubit it will "project" to either pure 0 or pure 1 state in the basis in which measurement has been carried out. To illustrate this let us consider circularly polarized light entering a polarizing beam splitter (PBS), Fig. 1. PBS decomposes polarization of incident light and sends linear horizontal component on one output port and linear vertical component to the other port.

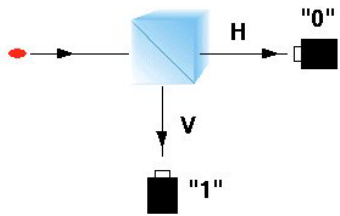

Fig. 1. Circularly polarized photon splits onto a linear horizontal/vertical analyzer with 50% chance to finish in either of the two output ports.

Thus, circularly polarized photon has equal content of both linear polarizations but since it cannot be split in half it has exactly 50% chance to exit either port. If now we label one of the ports as "0" and the other as "1" we immediately get a theoretically perfect RNG whose randomness is guaranteed by laws of quantum physics. Quantum RNG's (based on this or other principles) can be made quite good and imperfections (such as non-perfect circular polarization, beam splitter port axis misalignment, memory effects in photon detectors, etc.) of any type can be measured independently of the bit generation process so their effect on random numbers can be estimated with precision.

In conclusion, the most distinctive characteristic of quantum RNG is that it offers scientific proof of randomness. Careful practical applications came sufficiently close to theoretical idealization and allow for an independent assessment of imperfections which can, if required, be dealt with by information theoretic post-processing (see below). Quantum RNG's are the best choice for cryptography and other application which critically require true random numbers. Most significant drawback of present solutions is that they make use of bulky physical objects and therefore can not be miniaturized to the chip level using present technologies. Nascent science and technology of optical chips offers a promising avenue for Q-RNG's.

*D. Post processing*

There exist plethora of post processing algorithms whose purpose is to eliminate imperfections in "raw" random numbers produced by imperfect hardware generators. It is important to note that ad hoc, naive processing can lead to unexpected problems. For example it is usually considered a good idea to apply von Neumann de-biasing scheme in order to completely remove any bias from the sequence of bits. However, it is often overlooked that this works only if bits are completely independent (no correlations). The following extreme example illustrates how miserably von Neumann's procedure can fail. Let us consider the sequence: 101010101010... It obviously readily has no bias. After application of von Neumann de-biasing the sequence reads: 0000000.... which is a maximally biased sequence ! The reason for this un-expected result is that in the original sequence are maximally (anti-correlated). Generally, if the raw string is correlated, naïve de-biasing procedure may even increase the bias or create other unexpected statistical deficiencies.

More scrutinized approach is offered by the young theory of extractors [8]. A randomness extractor is an algorithm that converts a long weakly random sequence into a shorter sequence with almost perfect randomness. For some randomness sources IT provable extractors exist but no single randomness extractor currently exists that has been proven to work when applied blindly to any type of a high-entropy source.

Problem with extractor algorithms is that they require a memory buffer and a lot of CPU which slows down the available bit rate. In our opinion, a good physical RNG should be post-processing free.

*E. Randomness evaluation (testing)*

Most randomness tests check one or more statistical properties of long sequences of random numbers, for example bias, serial autocorrelation etc. Some compilation of tests are more oriented towards problems in PRNG's (eg. DIEHARD [9]) some more to hardware RNG's (eg. ENT [10]) while some are of general nature (eg. NIST STS [11]). The unfortunate fact is that there is an infinite number of statistical properties which truly random numbers must satisfy. Tests themselves are not perfect: some contain errors discovered latter [11], [12] or constants of questionable precision obtained by simulation using "trusted" random number generators.

The most important notion about statistical testing is the following: if a generator passes all known statistical tests it does not prove that it is random: it only means that it passes all *currently* known randomness tests. Tomorrow it can fail some new test or it already fails in the way known only to its constructors.

Nevertheless, randomness testing is important for constructors of RNG's. In some cases (especially for quantum RNG's) one can reasonably expect only certain type of imperfections and use tests sensitive to these problems only.

## III. RANDOM NUMBERS IN QUANTUM CRYPTOGRAPHY

Quantum cryptography is a protocol of public agreement of a symmetric cryptographic key, meaning if two parties A and B possess a small common secret key then using this protocol they will be able to establish a common secret key of any length. This cryptographic function is also known as "secret key growing". The ultimate goal of establishing a long secret key is to use it as a one time pad and thus obtain transfer of data in absolute secrecy. There are several mathematically identical QC protocols. The first one, named after its creators BB84 appeared in 1984 and has been experimentally realized in 1991 [1].

In the BB84 scenario, Alice and Bob are connected via two different channels: the quantum channel (usually well shielded optic fiber) capable of conducting single photons of light and an un-secured "classical" channel such as a telephone line, radio link or internet.

Here is the simplified schematic of how the protocol works.

Alice can prepare photons in different polarization states. In order to establish a secret key, Alice sends to Bob a sequence of *random numbers* encoded in photon polarizations as follows:

"1" is equi-probably encoded either as linear-vertical (LV) or left-circular (LC) polarization, while "0" equi-probably encoded either as linear-horizontal (LH) or right-circular (RC). Bob, has two polarization analyzers: one which can correctly measure linear polarizations (L) and the other which can correctly measure circular polarizations (C). Alice chooses one polarization *at random*, prepares the photon and sends it to Bob. Bob chooses one of the two analyzers *at random* and measures with it the photon received from Alice. If, by chance, Bob has chosen right polarizer he will receive 0 or 1 as sent by Alice. If Bob has chosen wrong polarizer he will receive 0 or 1 with equal probability regardless of what Alice has sent. So, after receiving a photon from Alice Bob announces (over authenticated but nit secret public channel) which polarizer he has just used (L or C). Note that this says nothing to potential eavesdropper about the value of the bit Bob has got. Alice responds with "Keep it !" or "Trash it !". So, bit by bit the two of them are building their secret key. Laws of quantum mechanics prevent qubit from being faithfully copied so an eavesdropper can obtain only a limited information about Alice's and Bob's string and furthermore eavesdropping can be detected by Alice and Bob.

It is straightforward to see that the whole protocol would be completely insecure if only the eavesdropper could calculate (or predict) either Alice's random numbers or Bob's random numbers or both. From analysis of the secret key rate presented in [13] it is obvious that any guessability of random numbers by the eavesdropper would leak relevant information to him, thus diminishing the effective key rate. It is intriguing (and obvious) that in the case that the eavesdropper could calculate the numbers exactly, the cryptographic potential of the BB84 protocol would be zero. This example shows that the local random number generators assumed in BB84 are essential for its security and should not be taken for granted.

Apart from what has been described above, the BB84 protocol has two more sub-protocols. Namely, due to the quantum incoherence, loses in the quantum channel or eavesdropping Alice and Bob will not have the exact same strings of bits after the first phase, although the two strings will have a lot of common information. Therefore the second sub-protocol, the „data reconciliation", is used to equalize the two strings, albeit at a cost of leaking some small information to an eavesdropper. Fortunately, Alice and Bob can calculate a lower limit of their mutual information after the two initial phases and then perform the privacy amplification phase in order to arrive to a shorter but much more private key. These two sub-protocols require further random numbers.

The protocol BB84 is considered information theoretically proven [14], [15] meaning that an attacker simply has no enough information to calculate the plaintext even given infinite computing resources. This is in strong contrast with widely used "deterministic cryptography" (see below for definition) where an attacker has enough information to calculate the key except that it would probably require insurmountably large computation resources and/or time. The caviat with QC is that the security proof holds only against family of attacks considered in the proof. Unfortunately, with time, it became evident that unexpected attacks on QC which utilize various quantum effects (such as energy-momentum co-entanglement in [16]) are feasible which makes QC much less "untouchable".

On top of that, as with any other cryptographic procedure, some problems in real-world implementation of the protocol, especially of the quantum channel and real photon detectors could be used to weaken the cryptographic security of the protocol and open pathways for attacks.

A beautiful demonstration of serious weakening and even 100 percent breaking of the key without any notice to legitimate parties has been made by Makarov et al. in 2010 [17], [18]. The demonstration has been made on the commercial QC system from Swiss company IdQuantique, based in Geneva, Switzerland, and one by MagiQ Technologies, based in Boston, Massachusetts. Improvements that would make QC resistant to those attacks are possible and have been proposed [19], but the lesson learned from this is that even protocols whose theoretical base is proven secure in some scenario are not to be automatically assumed immune to all practical attacks.

In conclusion, quantum cryptographic protocol BB84 requires that *both* Alice and Bob posses their *private* (local) provable random number generators. This is a highly critical requirement. Note that a public server of random numbers can not substitute for local generators because the random numbers would have to be delivered to Alice and Bob in perfect secrecy in the first place, and the server would have to be trusted.

## IV. RANDOM NUMBERS IN STATISTICAL CRYPTOGRAPHY

Statistical cryptography has been invented by U. Maurer in 1991. So called SKAPD protocol [20] resembles quantum cryptography and likewise consists of three sub-protocols. In fact the last two sub-protocols (the Data reconciliation and the Privacy amplification) are the same as in QC. However, the first sub-protocol, named "Advantage distillation" (AD) is completely different and it does not involve the quantum channel which potentially makes it more practical. Instead, it requires something called „binary channel with noise" which is theoretically a classical communication channel complemented with a provable RNG.

The condition for successful key agreement is that prior to AD protocol common information shared by Alice and Bob is greater that common information shared by either Alice and Eve or Bob and Eve.

The practical problem with SKAPD is that it contains an unspoken "zeroth phase" in which Alice and Bob obtain their partially correlated initial strings of bits which satisfy the above condition. There is no known plausible way to make the zeroth phase possible although some scenarios have been proposed (scanning surface of the Moon, listening to noise from far-away galaxies, taking big chunks of internet data, etc.).

## V. RANDOM NUMBERS IN DETERMINISTIC CRYPTOGRAPHY

What we call here "deterministic cryptography" is what is widely known as just "cryptography". It is the contemporary cryptography based on Galois groups, prime factoring or operations on elliptic curves and use of random data. Since all such security protocols are by definition deterministic and therefore reversible, the only true security resource is that nondeterministic part – a key or one-time data which is supposed to be "random". Quality and provability of randomness are therefore crucial for security of the whole system.

It is the fact that the deterministic cryptography is the only one in the wider use and that most cryptographers are not aware of or do not care of existence of either quantum cryptography or statistical cryptography because apparently they are not yet practical and/or trusted enough. Therefore it is important to understand what makes contemporary commercial-grade protocols secure and what could be done to get the maximum security out of them. Without ambition to make a comprehensive review here let us have a look at several examples supporting this conclusion.

1. Diffie-Hellman key establishing protocol [21]. This protocol enables the same functionality as the above mentioned BB84 and SKAPD protocols and is used for example in https secure internet protocol to establish a session key. Protocol requires from both parties (Alice and Bob) to generate private random data, and after some operations send them to each other. More resistant version of DH requires further random data used for digital signatures. A vulnerability of the PRNG built in an early version of Netscape internet browser led to complete compromise of the subsequent cryptographic protocol. An example is attack to the Netscape's 40 bit RC4-40 [22] challenge data and encryption keys, which was able to break https protocol in a minute or so, is described in [23]. The authors of that article suspect that 128 bit version RC4-128 would not be much harder to break either if seeding is done in a similar fashion.

2. RSA public key protocol relies on generation of public and private keys separately by Alice and Bob. In order to create a private/public pair of keys it is necessary to generate two unique, large prim numbers. Already calculating prim number candidates involves random numbers. After that, candidates need to be tested for primality using fast Miller-Rabin test which only works properly if fed by random numbers. Additional one-time random numbers may be used in the process of actual communication. Where high-entropy hardware random bits are not available or time-expensive (like on a typical PC computer) there is a tendency to "expand" a short random string to a long one by pseudo-random methods. This approach can create serious cryptographic weaknesses because an attacker must guess much smaller number of bits than he would in case of use of truly random numbers.

3. Similarly, a research of cryptographic attack on partially pseudo-random number generator of an AES based commercial cryptographic system is described in [24].

To conclude, in deterministic cryptography random numbers are the only part of the protocol which is different from point to point and furthermore their true randomness is sometimes a prerogative for correct calculations. Therefore, even though most of deterministic cryptography primitives are not IT secure, using true random numbers ensures highest achievable security with these methods.

## VI. CONCLUSION

In this survey we show the importance of random numbers for strength of cryptographic protocols not only for quantum and stochastic cryptographies where random numbers are an essential part of data exchanged between communicating parties but also for contemporary deterministic cryptography where un-guessability and maximal entropy of the random numbers used therein maximizes overall cryptographic strength.